\documentclass[aip,jcp,reprint]{revtex4-1} 

\usepackage{graphicx}
\usepackage{xcolor}
\usepackage{calc}

\usepackage{titlesec}
\titlespacing\section{0pt}{6pt plus 2pt minus 2pt}{6pt plus 2pt minus 2pt}
\titlespacing\subsection{0pt}{4pt plus 2pt minus 2pt}{4pt plus 2pt minus 2pt}

\usepackage[colorlinks=true,linkcolor=blue,urlcolor=blue,citecolor=blue]{hyperref}

\makeatletter
	\def\Dated@name{Published in \textit{The Journal of Chemical Physics} 148(24): 241401, 2018. \href{https://doi.org/10.1063/1.5043213}{DOI: 10.1063/1.5043213}}
\makeatother

\begin{document}

\title{Guest Editorial: Special Topic on Data-enabled Theoretical Chemistry}

\author{Matthias Rupp}
\email[]{matthias.rupp@fhi-berlin.mpg.de}
\homepage[]{http://www.mrupp.info}
\affiliation{\begin{minipage}{\linewidth}\small Fritz Haber Institute of the Max Planck Society, Faradayweg 4--6, 14195 Berlin, Germany\end{minipage}}

\author{O. Anatole von Lilienfeld}
\email[]{anatole.vonlilienfeld@unibas.ch}
\affiliation{\begin{minipage}[t]{0.8\linewidth}\small Department of Chemistry, Institute of Physical Chemistry and National Center for Computational Design and Discovery of Novel Materials, University of Basel, 4056 Basel, Switzerland\end{minipage}}

\author{Kieron Burke}
\email[]{kieron@uci.edu}
\affiliation{\begin{minipage}[t]{0.8\linewidth}\small Departments of Chemistry and of Physics, University of California, Irvine, CA 92697, USA\end{minipage}}

\date{}

\begin{abstract}
A survey of the contributions to the Special Topic on Data-enabled Theoretical Chemistry is given, including a glossary of relevant machine learning terms.
\end{abstract}

\maketitle


\nocite{g2015,jm2015,m1997,r2015,rn2009,ge2003b,sv2010,s2010b,rbpmk2017,bhsstv2007,lv2007,s2012b,ss2002,hss2008,bgv1992,ssm1998,ssm1998,bgv1992,htf2009,rw2006,b1996,mom2012,gbc2016,p1901,j2004b}
\nocite{cjsvd1994,mfmsa1996,v2001,mfmsa1996,m2017b,ke2004,l2018}
\nocite{Yang,Eickenberg,Faber,Lubbers,Gastegger,Gubaev,Collins,Bereau,Imbalzano,Fujikake,Smith,Unke,Schutt,Pasquale,Wood,Zeni,Herr,Zhang,Quaranta,Qu,Nguyen,Kamath,Artrith,Schmidt}
\nocite{Li,Kiyohara,Seko,Dieb,Mardirossian,prtssc2005,Nagai,lz2014,Ji,Hollingsworth,Nagai,Seino,Boninsegna,Wehmeyer,Matsunaga,Pernot,Jorgensen,Afzal,Pilania,Schmitz,Sorensen,Botlani,Cipcigan}


\noindent
\textbf{NOMENCLATURE}

	\bigskip

	\noindent\begin{tabular}{@{}p{5em}p{\linewidth-2\tabcolsep-5em}@{}}
		AI      & Artificial Intelligence, see Sec.~\ref{secFields} \\
		B3LYP   & Becke, three-parameter, Lee-Yang-Parr, a hybrid DFT functional \\
		CCSD(T) & Coupled Cluster with Single, Double and perturbative Triple excitations, an electronic structure method  \\
		DFT     & Density Functional Theory, an electronic structure method \\
		DFTB    & Density Functional Theory Tight Binding, an electronic structure method \\
		DNN     & Deep Neural Network, see Sec.~\ref{secAlgs} \\
		EAM     & Embedded Atom Model/Method, an interatomic potential \\
		GAP     & Gaussian Approximation Potential, a machine learning potential \\
		HOMO    & Highest Occupied Molecular Orbital \\
		KRR     & Kernel Ridge regression, see Sec.~\ref{secAlgs} \\
		LUMO    & Lowest Unoccupied Molecular Orbital \\
		MAE     & Mean Absolute Error, see Sec.~\ref{secModelBuilding} \\
		MD      & Molecular Dynamics, a simulation technique \\
		ML      & Machine Learning, see Sec.~\ref{secFields} \\
		MP2     & M{\o}ller-Plesset perturbation theory to Second order, an electronic structure method \\
        QM/MM   & Quantum Mechanics/Molecular Mechanics, a molecular simulation method \\
		(A)NN   & (Artificial) Neural Network, see Sec.~\ref{secAlgs} \\
		QSPR    & Quantitative Structure-Property Relationship, see Sec.~\ref{secFields} \\
		RMSE    & Root Mean Squared Error, see Sec.~\ref{secModelBuilding} \\
		SINDy   & Sparse Identification of Nonlinear Dynamics, a machine learning method \\
		SNAP    & Spectral Neighbor Analysis Potential, a machine learning potential \\
		SVM     & Support Vector Machine, see Sec.~\ref{secAlgs} \\
		tICA    & time structure Independent Component Analysis, see Sec.~\ref{secAlgs} \\
	\end{tabular}

\vfill

\pagebreak

\section{Introduction}

Welcome to the Journal of Chemical Physics Special Topic on data-enabled theoretical chemistry.  
We expect that this will be a timely addition to this new and rapidly evolving field,
with a variety of articles from the front lines.

Unless you have disconnected from all social media, you will have noticed that
artificial intelligence, machine learning, big data, and other vague
but computer-driven terms have invaded many realms of public life.
Facial recognition software has been revolutionized by machine learning,
cars now drive themselves,
the world's best chess and go players are algorithms, and perhaps someday soon
they'll even be able to recommend a good movie.

The same revolution has also been occurring in many branches of
theoretical and computational chemistry, driven by the same force:
the never-ending increase in data being generated by computers.
Our Special Topic is devoted to data-enabled chemistry, which we interpret broadly.  We cover essentially all algorithmic developments
that fit under the broad rubric of machine learning, using varying amounts
of data, and driven by applications from small molecule chemistry to
materials science to protein behavior.

\begin{figure}[htb]
	\includegraphics{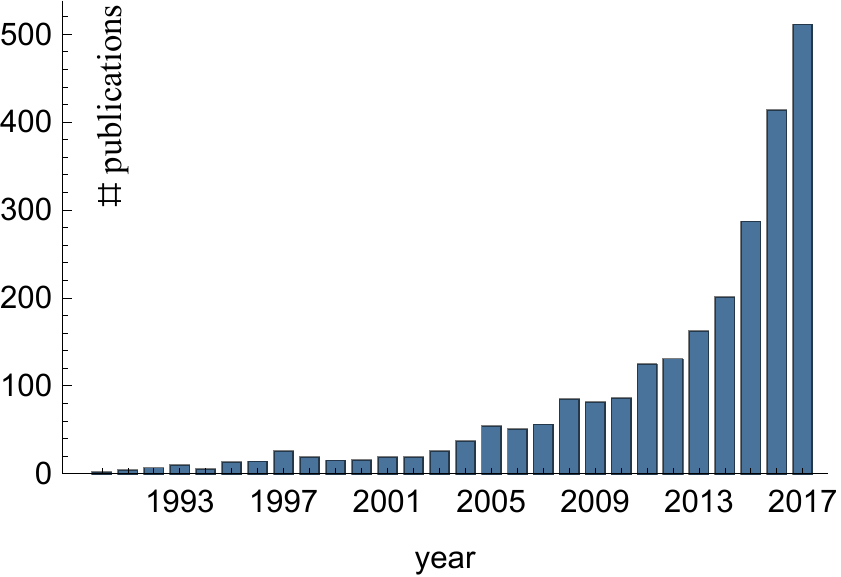}
	\caption{\emph{Number of publications per year} from a web of science search for articles with topics of machine learning and either chemistry or materials, taken June~5, 2018. The average number of citations per article is 12.\label{cites}}
\end{figure}

\newcommand{\TBa}[1]{\begin{minipage}[c][2.5em]{33ex}\raggedright #1\end{minipage}}
\newcommand{\TBb}[1]{\begin{minipage}[c][2.5em]{35.5ex}\raggedright #1\end{minipage}}
\newcommand{\TBc}[1]{\begin{minipage}[c][2.5em]{26.5ex}\raggedright #1\end{minipage}}

\begin{table*}[p]
	\vspace*{-2.5\baselineskip}
	\caption{\emph{Overview} of contributions to the Special Topic.\hfill\mbox{}}\label{tabOverview}

	\smallskip\small

	\begin{ruledtabular}
	\begin{tabular}{@{}llp{26.5ex}lp{33ex}p{35.5ex}@{}}
		Ref. & Sec. & ML Method & QM Method & Systems & Keywords \\ \hline
		\citenum{Yang}          & \ref{secProperties}  & NN                                        & DFT                & Hydrocarbon molecules             & Size-independence                          \\
		\citenum{Eickenberg}    & \ref{secProperties}  & Multilinear regression                    & DFT                & Small organic molecules           & Representation, wavelets                   \\
		\citenum{Faber}         & \ref{secProperties}  & KRR                                       & DFT                & Organic molecules, water, solids  & Representation, many-body terms            \\
		\citenum{Lubbers}       & \ref{secProperties}  & NN                                        & DFT                & Small organic molecules           & NN architecture                            \\ 
		\citenum{Gastegger}     & \ref{secProperties}  & NN                                        & DFT                & Small organic molecules           & Representation, symmetry functions         \\ 
		\citenum{Gubaev}        & \ref{secProperties}  & Regression                                & DFT                & Small organic molecules           & Polynomial fit, active learning            \\ 
		\citenum{Collins}       & \ref{secProperties}  & KRR                                       & DFT                & Small organic molecules           & Graph-based representation                 \\ 
        \citenum{Hy}            & \ref{secProperties}  & NN                                        & DFT                & Organic molecules                 & Covariant compositional networks           \\ 

		\citenum{Bereau}        & \ref{secPotentials}  & KRR                                       & DFT, CCSD(T)       & \TBa{Dimers, hydrogen-bonded complexes, and others} & Non-covalent interactions \\
		\citenum{Imbalzano}     & \ref{secPotentials}  & GPR, NN                                   & DFT                & \TBa{Liquid water, Al-Si-Mg alloy, organic molecules} & Feature selection      \\
		\citenum{Fujikake}      & \ref{secPotentials}  & GPR                                       & DFT                & Li-C guest-host systems           & Combination of potentials                  \\
		\citenum{Smith}         & \ref{secPotentials}  & NN                                        & DFT                & Small organic molecules           & Active learning                            \\
		\citenum{Unke}          & \ref{secPotentials}  & NN                                        & DFT                & Small organic molecules           & Molecular properties                       \\
        \citenum{Schutt}        & \ref{secPotentials}  & DNN                                       & DFT                & \TBa{Organic molecules, bulk crystals, C$_{20}$-fullerene} & DNN architecture  \\
		\citenum{Pasquale}      & \ref{secPotentials}  & GPR                                       & DFT, force field   & Na$^+$, Cl$^-$ ion-water clusters & Ion-water interactions                     \\
		\citenum{Wood}          & \ref{secPotentials}  & \TBc{Regularized linear regression}       & DFT                & Tantalum                          & Bispectrum quadratic terms                 \\ 
		\citenum{Zeni}          & \ref{secPotentials}  & GPR                                       & DFT                & Ni nanoclusters                   & Interatomic forces, $k$-body kernels       \\ 
		\citenum{Herr}          & \ref{secPotentials}  & NN                                        & DFT                & Nicotine, water cluster           & Sampling, meta-dynamics                    \\ 
        \citenum{Zhang}         & \ref{secPotentials}  & NN                                        & DFT                & Cu surface grain boundaries       & Hybrid QM/ML models  \\

		\citenum{Quaranta}      & \ref{secSpecific}    & NN                                        & DFT                & Water/ZnO(10$\bar{1}$0) interface & Anharmonic vibrational spectra \\ 
		\citenum{Qu}            & \ref{secSpecific}    & Linear regression                         & CCSD(T)            & Formic acid dimer                 & Dipole moment surface, infrared spectrum   \\ 
		\citenum{Nguyen}        & \ref{secSpecific}    & NN, GPR                                   & CCSD(T)            & Water (ice, liquid, clusters)     & Representation, invariant polynomials      \\ 
		\citenum{Kamath}        & \ref{secSpecific}    & NN, GPR                                   & Force field        & Formaldehyde                      & Comparison, vibrational spectra \\ 

		\citenum{Artrith}       & \ref{secStability}   & NN, genetic algorithm                     & DFT                & Li$_x$Si alloys                   & \TBb{Phase diagrams of amorphous materials} \\ 
		\citenum{Schmidt}       & \ref{secStability}   & Regression trees                          & DFT                & AB$_2$C$_2$ ternary intermetallics& Stable compound search                     \\ 
		\citenum{Li}            & \ref{secStability}   & Clustering                                & Harris approximation & Rigid-molecule crystals         & Crystal structure prediction               \\ 
		\citenum{Kiyohara}      & \ref{secStability}   & Monte Carlo tree search                   & EAM                & Ag, Co grain boundaries           & Segregation                                \\ 
		\citenum{Seko}          & \ref{secStability}   & Binary classification trees               & DFT                & Inorganic crystals                & Recommender system                         \\ 
		\citenum{Dieb}          & \ref{secStability}   & \TBc{Monte Carlo tree search, GPR}        & DFT                & Boron-doped graphene              & Stable structure search                    \\ 

		\citenum{Mardirossian}  & \ref{secFunctionals} & \TBc{Subset selection, outlier detection} & DFT                & Main group chemistry              & Doubly hybrid functional                   \\ 
		\citenum{Nagai}         & \ref{secFunctionals} & NN                                        & DFT                & Model systems                     & \TBb{Hartree-exchange-correlation potential}\\
		\citenum{Ji}            & \ref{secFunctionals} & KRR                                       & DFT                & Organic molecules                 & Representation                             \\ 
		\citenum{Hollingsworth} & \ref{secFunctionals} & KRR                                       & DFT                & Model systems                     & Exact conditions                           \\ 
		\citenum{Seino}         & \ref{secFunctionals} & NN                                        & DFT                & Atoms and molecules               & Kinetic energy density functional          \\ 

		\citenum{Boninsegna}    & \ref{secAnalysis}    & Sparse regression                         & Analytic potential & Model systems                     & Stochastic dynamical equations             \\ 
		\citenum{Wehmeyer}      & \ref{secAnalysis}    & Time-lagged autoencoder                   & Force field        & Model systems, villin peptide     & \TBb{Slow dynamics, dimensionality reduction} \\
		\citenum{Matsunaga}     & \ref{secAnalysis}    & Markov state model,tICA                   & Force field        & Dye-labeled polyproline-20        & Dynamics, transition probabilities         \\ 

		\citenum{Pernot}        & \ref{secOther}       & None                                      & DFT                & Various (G3/99 test set)          & Error statistics                           \\ 
		\citenum{Jorgensen}     & \ref{secOther}       & Autoencoder, NN                           & DFT                & Donor-acceptor polymers           & Screening, solar cells                     \\ 
		\citenum{Afzal}         & \ref{secOther}       & SVM                                       & DFT                & Organic polymers                  & Refraction index                           \\ 
		\citenum{Pilania}       & \ref{secOther}       & KRR                                       & DFT                & \TBa{Perovskite oxides, elpasolite halides} & Lanthanide-doped scintillators \\
		\citenum{Schmitz}       & \ref{secOther}       & GPR                                       & CCSD(T)            & Small organic molecules           & Geometry optimization                      \\ 
		\citenum{Sorensen}      & \ref{secOther}       & Clustering                                & DFTB               & Anatase TiO$_2$(001)              & Global structure optimization              \\ 
		\citenum{Botlani}       & \ref{secOther}       & SVM, graph analysis                       & Force field        & Tyrosine phosphatase 1E           & Proteins, dynamic allostery                \\ 
		\citenum{Cipcigan}      & \ref{secOther}       & Data analysis                             & Force field        & Antimicrobial peptides            & Visualization                              \\

	\end{tabular}
	\end{ruledtabular}
\end{table*}

\pagebreak

\enlargethispage*{1.4\baselineskip}

In Fig.~\ref{cites}, we show papers being published involving machine learning and chemistry or materials over the last three decades.
The absolute rate is rather arbitrary, depending on the precise search terms, but the rapid growth is robust, as is the average citation rate of each article.  
There is no doubt that data-enabled chemistry is rapidly making a large impact in the field.

This editorial is designed for non-experts who are outside this field, and
trying to figure out what is going on and how they might want to get in
on the action.
We provide a brief glossary of machine-learning terms for non-experts in Sec.~\ref{secGlossary}, focusing on the concepts and algorithms used most often in physical chemistry and materials science.
In Sec.~\ref{secContributions}, using the introduced terminology, we briefly survey the contributions in this Special Topic, grouped by the physical and chemical processes and systems to which they are applied.

A nomenclature and a table are provided to aid the reader:
The Nomenclature summarizes the used abbreviations and Table~\ref{tabOverview} presents an overview of all articles in the Special Topic,
acting as a quick guide to the methods (both quantum chemical and computer science) and the systems included.
Not only is it a quick way to find something in the issue, but it also represents a snapshot of the state of the field today.


\section{SOME DATA-ENABLED TERMINOLOGY}\label{secGlossary}

\newcommand{\glos}[1]{{\textsc{#1}}}

This section is an introduction to common terminology in machine learning, with an emphasis on those concepts currently in use in the applications in this Special Topic.  
Terms used both in this editorial and throughout the Special Topic are set in small capitals, followed by their explanation.  
This is by no means a comprehensive explanation, and interested readers should consult further sources for more detailed explanations.  

\subsection{Machine learning and related scientific fields}\label{secFields}

\begin{figure}
	\includegraphics[width=\linewidth]{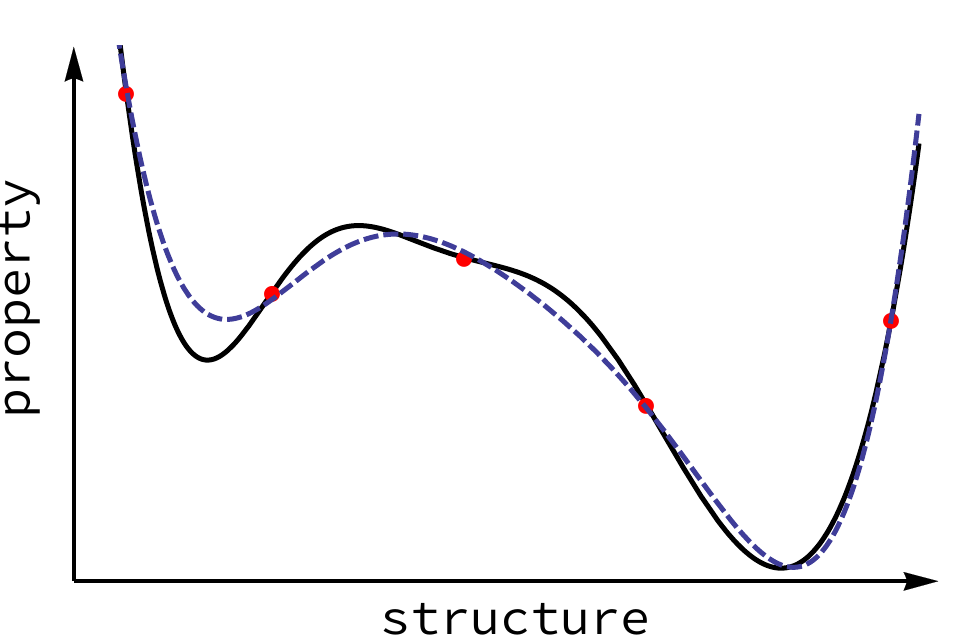}
	\caption{\emph{Sketch illustrating the idea of machine learning}, \cite{r2015} using prediction of molecular energies as an example.
		The horizontal axis represents molecular space (molecules are points on the axis), the vertical axis represents energy.
		Instead of calculating all energies (solid line), only a few reference calculations are done (dots),
		and machine learning is used to learn the mapping from molecule to energy (dashed line).}\label{figSketch}
\end{figure}

\glos{Machine learning} (\glos{ML})~\cite{g2015,jm2015} is an umbrella term referring to algorithms that improve with data ("learn from experience"),~\cite{m1997} mostly for analysis or prediction.
Instead of being explicitly programmed to solve a specific problem, these algorithms rely on given data to make statements about new data.
An example for a ML algorithm is regression (Fig.~\ref{figSketch}):
Based on a finite number of points (\glos{examples}, \glos{samples}), a function is inferred which enables predictions for new examples;
the fit gets better the more examples there are.
While ML encompasses many different tasks besides regression,
such as classification, dimensionality reduction, clustering, anomaly detection, optimization,
and offers a wide variety of specific algorithms,
such as Gaussian process regression, support vector machines, principal component analysis, (deep) neural networks,
the underlying principle of data-driven improvement remains the same.

\glos{ML} is related to, but distinct from, artificial intelligence and data mining.
\glos{Artificial intelligence} (\glos{AI})~\cite{rn2009} is the study of machines that exhibit intelligent behavior.
The scope of this field is less clear-cut, evidenced by the lack of a formal definition of intelligence.
\glos{AI} traditionally involves (symbolic) knowledge representation and logical reasoning.
\glos{Data mining} is similar to ML, but more concerned with extraction of new patterns in large datasets.
\glos{Pattern recognition} is essentially a synonym for ML.
For the more recent term \glos{data science}, no consensus has emerged yet, but it is often used to mean applied \glos{ML} and statistics.

Two major application areas of \glos{ML} closely related to this Special Topic are cheminformatics and materials informatics.
\glos{Cheminformatics}\cite{ge2003b} (also chemoinformatics) is at the intersection of chemistry and computer science.
In particular, \glos{quantitative structure-property relationships}
(QSPR)~\cite{sv2010} relate molecular features or descriptors to, usually experimental, molecular properties,
and \glos{virtual screening} \cite{s2010b} is the computational screening of large databases for compounds with desired properties.
\glos{Materials informatics} \cite{rbpmk2017} is a newer field at the intersection of materials science and computer science.

\subsection{Types of problems machine learning addresses}

One way to categorize problem types in ML is according to the type of examples involved.
In \glos{supervised learning}, examples are pairs of input~$x$ and label~$y$, for example molecules and their energy, and the task is to predict the label of new examples, that is to learn the function $f : x \rightarrow y$.
In \glos{unsupervised learning}, only inputs~$x$ are given, and the task is to find structure in the data.
An example would be identifying a reaction coordinate from molecular dynamics (MD) data.
Mixed forms are possible as well: In \glos{semi-supervised learning}, only some examples are labeled, the idea being that large amounts of unlabeled data can still help with predictions by characterizing the manifold the data lie on.
An example would be a large combinatorial chemistry database of molecules where only some have been measured or calculated.

Frequent types of problems within supervised learning are classification and regression.
In \glos{classification}, labels belong to a finite set of outcomes, where one distinguishes between two possible labels in \glos{binary classification}, for example active and inactive, and, multiple possible labels in \glos{multi-class classification}, for example different phases.
The special case with only one possible label is \glos{one-class learning} (also \glos{novelty detection}, \glos{outlier detection}, or \glos{anomaly detection}), where examples from a single class are given and the task is to detect whether new examples fall outside of this class or not.
In \glos{regression}, labels are continuous.
Usually, these are scalar values, but vectors, distributions or other structured objects like graphs can also be predicted using \glos{structured-output learning}~\cite{bhsstv2007}.

Frequent problem types within unsupervised learning are dimensionality reduction and clustering.
In \glos{dimensionality reduction}~\cite{lv2007} the goal is to find a subspace or manifold of low dimension on which the data live.
\glos{Clustering} attempts to group samples into clusters such that samples within a cluster are more similar to each other than to samples in other clusters.

There are many other concepts that have found their way into data-enabled theoretical chemistry and materials science:
In \glos{active learning}~\cite{s2012b}, the training data are not sampled randomly but ``actively'' chosen by the \glos{ML} algorithm; this often enables achieving the same prediction error with much smaller training sets.
In \glos{reinforcement learning}, the \glos{ML} algorithm chooses an action from a set of possible actions based on the state of its environment.
It is then rewarded accordingly and the process repeats.
The goal of the algorithm is to maximize reward.

\subsection{Specific algorithms}\label{secAlgs}

Many ML algorithms exist, but the ones used most often in cheminformatics and materials informatics belong to two large families, kernel-based ML and (deep) artificial neural networks.

In \glos{kernel-based ML}, \cite{ss2002,hss2008} inputs $x$ are non-linearly transformed into a higher-dimensional space,
where problems can become linear with the right transformation.
As working directly in these high-dimensional feature spaces is impractical, kernel functions $k$ are used.
These are computed in the original input space, but yield inner product values, and thus geometric information, in the high-dimensional space.
Since their invention in the 1990s, \cite{bgv1992,ssm1998} many linear ML algorithms have been ``kernelized''.
Popular algorithms include \glos{support vector machines} (SVM), \glos{kernel principal component analysis}~\cite{ssm1998}, \cite{bgv1992} \glos{kernel ridge regression} (KRR),~\cite{htf2009} and \glos{Gaussian process regression} (GPR)~\cite{rw2006}  (also called \glos{Kriging} due to its origins in geostatistics).
While KRR is a frequentist algorithm and GPR is a Bayesian one, their predictions are formally identical, which is why the terms KRR and GPR are occasionally used interchangeably in practice.

Artificial \glos{neural networks} (NNs) \cite{b1996,mom2012} are repeated compositions of simple functions, where the input of one function are the weighted outputs of other functions.
These functions are typically arranged in consecutive layers.
In graph representations of NNs, vertices correspond to functions and edges correspond to weighted connections between them.
Determining the weights is a non-convex optimization problem.
\glos{Deep NNs} (DNNs) \cite{gbc2016} are characterized by having many layers of functions.
This depth enables them to learn internal representations of the data of increasing complexity and abstraction.

Kernel learning and NN are simply two different ways of fitting a flexible function to data.
Many other learning algorithms exist, including tree-based algorithms such as \glos{decision trees}, \glos{regression trees}, and \glos{random forests}.

A classic algorithm for dimensionality reduction is \glos{principal component analysis} (PCA),~\cite{p1901,j2004b} which finds orthogonal directions of maximal variance in the data.
Many variants of this idea exist, such as Independent Component Analysis (ICA), which finds independent latent variables and explains data as mixtures of these variables.
For \glos{time-structure Independent Component Analysis} (tICA), these variables are chosen to maximize autocorrelation.
A NN approach to dimensionality reduction are \glos{autoencoder} networks, where the size of function layers first decreases, then increases again and the task is to reproduce the inputs.
Having the data go through a ``bottleneck'' forces the autoencoder NN to find a low-dimensional representation of the data.

\subsection{Model building}\label{secModelBuilding}

Unlike classical potentials, which are parametrized once for a class of molecules or materials and then deployed, ML models, being more flexible mathematical functions, should be applied only to molecules or materials sampled from the same distribution as the ones used to train the model---otherwise, the ML model will operate outside of its \glos{domain of applicability}, resulting in uncontrolled and essentially arbitrary errors.
For this reason, ML models are often retrained, for example dynamically by adding training data ``on-the-fly'' during the course of a simulation.
Deciding when to make a prediction and when to do a reference calculation to update the model requires \glos{uncertainty estimates}, that is, assessments of the reliability of individual predictions.

The \glos{root mean squared error} (RMSE) 
is the canonical measure of how wrong a set of predictions is.
It is the RMSE that is minimized by many algorithms by default.
This typically leads to ``full'' solutions, such as all coefficients in an expansion being non-zero.
By contrast, \glos{sparsity} of solutions, that is, solutions with most coefficients zero, can be achieved by minimizing the \glos{mean absolute error} (MAE) 
or $L^1$-norm instead.

For \glos{validation} of a ML model, the errors reported must always be on \glos{out-of-sample} data, that is, data not used for training the model, including any pre-processing steps.
An easy way to achieve this is to set aside a \glos{hold-out set} in the beginning, to be used only for validation, and only after the ML model's training is complete.
For small datasets, where this might not be feasible, statistical validation techniques such as cross-validation can be used.
These essentially reuse the data by splitting it multiple times into a training and a hold-out set, then average over the results.

The training or model-building process can include steps such as optimizing free parameters, often called \glos{hyperparameters}, or, \glos{feature selection}, where only some of the descriptors or variables used to represent the inputs are retained.
\glos{Hyperparameter optimization} usually is a non-convex optimization problem, but well-behaved in practice.
For few parameters, it can be addressed via grid search, minimizing the hold-out RMSE over a logarithmic grid; alternatives are maximizing the likelihood of the model given the data, or choosing good values via heuristics.

The out-of-sample error of ML models must decay with training set size (otherwise it would not be machine {\em learning}). 
For many models, the leading error term varies as $a/N^b$, where $N$ is number of training data.\cite{cjsvd1994,mfmsa1996,v2001}
\glos{Learning curves} are plots of the out-of-sample prediction error as a function of~$N$, usually on a log-log scale.


\vspace*{2pt}

\enlargethispage*{0.75\baselineskip}
\section{SURVEY OF AREAS COVERED}\label{secContributions}

We next survey the areas covered by the articles in our Special Topic.  
We have organized them according to the type of chemical problem being addressed, as far as is possible.  
This makes it easier to see both the breadth of the problems and which topics have the most interest, as well as to compare different \glos{ML} approaches to the same problem.

\subsection{Prediction of energies and other properties throughout chemical compound space}\label{secProperties}

Chemical space is astronomically vast.~\cite{m2017b,ke2004}
Given some molecule, defined by its number of electrons and
the set of nuclei at their equilibrium geometries, we can typically
predict its observables with satisfying accuracy using ab initio quantum chemical methods such as
CCSD(T) in a sufficiently large basis.  This is feasible for smaller molecules, and DFT can be used (less reliably) for larger ones.
But even DFT (or computationally less demanding semi-empirical quantum chemistry methods)
is not fast enough to search all of chemical compound space, whose size grows combinatorially with the number of atoms and distinct elements.  Thus, an important problem is to search chemical compound space to find new drugs (and materials space to find new materials) with desired functionalities.

A basic property is the ground-state energy of a molecule.  
But there are also many other interesting properties at the ground-state configuration,
such as dipole moments, ionization potentials, and vibrational frequencies.
Some of these can be extracted from the same electronic structure calculation
from which the molecule's energy was obtained, while others
require additional computation.  Given the
impossibility of calculating all properties of all possible molecules,
it is interesting to ask if a ML algorithm, trained on known examples, can be used to predict the properties of
new molecules at much reduced computational cost.~\cite{l2018}
If so, chemical compound space can be searched orders of magnitude more quickly.  
Many groups are therefore formulating ways to do this.

Note that often researchers use DFT (or even DFTB) results for both training and testing their algorithms.  In those cases, the ML algorithm is tested against the DFT calculations, not experiments or more accurate quantum chemical methods.  The idea is that, once an algorithm is sufficiently robust and useful, it can then be trained on more accurate data and, presumably, work just as well.  These days, many ML approaches already produce MAE below those typical of density functionals.

Yang et~al.~\cite{Yang} introduce a size-independent \glos{NN} model of heats of formation trained on small organic molecules that can be applied to large molecules.
For these, the MAE from reference B3LYP numbers is reduced to 1.7\,kcal/mol.

On the other hand, Eickenberg et~al.~\cite{Eickenberg} introduce a ML model based on a solid harmonic wavelet scattering representation of organic molecules and demonstrate competitive performance for predicted atomization energies.  
Meanwhile, Hy et~al.~\cite{Hy} use a new kind of \glos{NN}, called a covariant compositional network, to deduce properties from molecular graphs alone, yielding promising results on databases of small molecules.

Often, the efficiency of a ML algorithm depends crucially on the way the data are represented. 
Faber et~al.~\cite{Faber} introduce a many-body representation of atoms in their environment 
and report ``chemical accuracy'' (1\,kcal/mol) for energies of organic molecules and solids with few thousand training points. 
Interpolation across the periodic table even enables prediction of energies of molecules with elements that were not included in the training set.

Lubbers et~al.~\cite{Lubbers} introduce a hierarchical NN approach with competitive performance for predicting atomization energies of organic molecules, as well as energies and forces of thousands of snapshots of benzene, malonaldehyde, salicyclic acid, and toluene.  Their method can also be applied to MD simulations and gives a measure of model uncertainty automatically.
Gastegger et~al.~\cite{Gastegger} develop element-specific weighting functions for atom-centered symmetry function-based representations in NNs.
Upon use of the weighting functions, they show that less symmetry functions are necessary and the prediction error of atomization energies in organic molecules is systematically reduced.

Gubaev et~al.~\cite{Gubaev} conceive a local tensor based ML approach which depends on the property being intensive or extensive, and they combine it with \glos{active learning} in order to achieve state-of-the-art performance for atomization energies, polarizabilities, and HOMO/LUMO eigenvalues in organic molecules.
Collins et~al.~\cite{Collins} show that graph-based molecular representations profit from inclusion of interatomic distance information while remaining size-independent, as evinced for competitive prediction errors of atomization energies in organic molecules.

\subsection{Interatomic potentials}\label{secPotentials}
\enlargethispage{-\baselineskip}

Classical MD simulations with interatomic potentials can
handle a million atoms or more and are used to study dynamic processes
in biology and chemistry. Unfortunately, the necessary computational efficiency is sometimes
obtained only at the expense of predictive power.
Typically, relying on complex classical force fields, which ignore
the underlying electronic structure and dynamics, can produce
inconsistent answers to important questions.
This limitation becomes especially acute when covalent bonds are formed or broken, when
atoms vary their hybridization state, or during considerable changes in chemical
environments, as in, for example, molten alloys.  
Then developing and testing force fields for all possible configurations become an unsurmountable task.
Given this challenge, and the relevance of a dynamic description of atomistic
processes throughout the exact sciences, 
a large number of articles in the Special Topic are devoted to the question 
if and how interatomic potentials can be constructed via ML, for example by
training on (usually) DFT calculations.

Bereau et~al.~\cite{Bereau} predict parameters for intermolecular force fields throughout chemical space. 
These parameters include atomic charges, dipole moments, quadrupole moments, polarizabilities, atomic electron density screening factors, and normalization constants.
Out-of-sample predictions on well established van der Waals benchmark datasets indicate errors below or about 1\,kcal/mol.

A crucial consideration for ML methods is the way in which the inputs are represented, which can have a strong impact on performance.
Imbalzano et~al.~\cite{Imbalzano} provide an automated protocol for \glos{feature selection}, showing how this can simplify construction of ML potentials.
They illustrate their procedure on \glos{NN} potentials for water and aluminum ternary alloys, as well as a \glos{GPR} potential for formation energies of molecules.

Gaussian approximation potentials (GAPs) are one of the success stories of ML in chemistry.
They provide an automated approach to constructing accurate interatomic potentials that recreate the underlying electronic structure energetics at a fraction of the computational cost.
Fujikake et~al.~\cite{Fujikake} study the issue of guest atoms in host structures, with the specific case of Li in C, showing how to add the Li interactions to a pre-existing GAP potential for C.

An important question, usually left to human bias and intuition, is the selection of data upon which to train:
When generating an interatomic potential, which sets of electronic structure calculations do you perform to create the database to train on and test against?  
Smith et~al.~\cite{Smith} present a fully automatic way of generating datasets for the specific purpose of training ML potentials.  Query-by-committee \glos{active learning} uses disagreements between predictions of different models to improve sampling and reduce the amount of data needed over random sampling.  Results are given on a new COMP6 database of small organic molecules containing CHNO.

Unke and Meuwly~\cite{Unke} are focused on creating methods that span both configurational space and chemical space.
Their method decomposes energy into local atomic contributions, with prediction errors on atomization energies on the order of half a kcal/mol after training on 35\,000 organic molecules.
They demonstrate predictive capability on both reactive and non-reactive MD simulations.

Advanced deep learning methods are applied by Sch\"utt et~al.~\cite{Schutt}
They present SchNet, a DNN that learns chemically relevant information about atom types across the periodic table.  
It is general and flexible and uses deep learning to avoid the need for clever choices of descriptors.  
It can be applied to both molecules and materials, and has been shown to reduce the computational cost of DFT-MD simulations of fullerenes by 3--4 orders of magnitude.

Another type of ML method is \glos{GPR} or \glos{Kriging}, and Di~Pasquale et~al.~\cite{Pasquale} use it to predict energies of ions solvated in water.  
Energies are based on atomic energies obtained from the topological partitioning called interacting quantum atoms.  
This method provides accurate results and is part of an advanced force field development, FFLUX.

Spectral Neighbor Analysis Potentials (SNAPs) express the energy of an atom linearly in terms of bispectrum components of neighboring atoms.
Wood and Thompson\cite{Wood} show that accuracy can be improved by including quadratic contributions at a modest increase in cost,
making it particularly suitable for large-scale \glos{MD} simulations of materials.

Metallic nanoclusters are important in many areas of chemistry, but realistic simulations are limited by the computational cost of DFT-MD.  
Zeni et~al.~\cite{Zeni} study such systems via classical $n$-body potentials derived from ML (``M-FFs'') by constructing $n$-body kernels that can be exactly mapped to non-parametric classical potential forms such as 3D splines.
This circumvents summing over training set entries for predictions, accelerating simulations by orders of magnitude.
They find that 2-body potentials are insufficiently accurate to capture the behavior of Ni clusters, but 3-body potentials are. 
Choice of training data also plays an essential role.

Another important question is which regions of configuration space to sample when constructing a ML force field.
Herr et~al.~\cite{Herr} explore application of metadynamics to training sets prior to selection for training.
Metadynamics avoids the problem of being stuck in the vicinity of local minima. 
In comparison to data retrieved from MD or normal-mode analysis based sampling, the resulting NN exhibits improved or more efficient performance.

Finally, QM/MM schemes are popular in computational molecular biology, but often suffer from limitations of the MM model and ambiguities at the interface.  
Zhang et~al.~\cite{Zhang} review this field for the specific case of a ML force field for the MM contribution.  
They point out both advantages and disadvantages of the ML approach.

\subsection{Potential energy surfaces of specific molecules}\label{secSpecific}

This section could arguably be part of the previous one.  
But in this section, the molecule is fixed, and a highly accurate potential energy surface is desired, for a fixed number of atoms.

A difficult problem is the simulation of water on oxide surfaces, as measured by infrared spectroscopy of OH anharmonic stretches.
MD simulations at the DFT level should be sufficiently accurate, but are too expensive computationally.
Quaranta et~al.~\cite{Quaranta} use a NN potential trained on such calculations to perform MD
and solve the nuclear Schr{\"o}dinger equation for a large number of configurations to determine vibrational spectra.
They find that many different species contribute in overlapping regions of the spectrum and that the stretching frequencies depend strongly on the hydrogen bonding.

For many purposes DFT-level calculations suffice, but not for the infrared spectrum of weakly bound dimers.
The potential energy surface is a function of all 45 internuclear distances and must be calculated at CCSD(T) levels of accuracy in order to accurately reflect the anharmonic couplings.
Qu and Bowman\cite{Qu} present a novel fit to the dipole moment and solve the nuclear Schr{\"o}dinger equation using various levels of anharmonic theory to generate the infrared spectrum.

Nguyen et~al.~\cite{Nguyen} perform a careful study of the general methodology for constructing interatomic potentials,
focusing on two- and three-body interactions in water using coupled-cluster energies. They compare different approaches:
\glos{GAP}, \glos{NNs}, and permutation-invariant polynomials, finding comparable levels of accuracy in the fit.

In a related way, Kamath et~al.~\cite{Kamath} study the potential energy surface of formaldehyde, in order to compare NNs with GPR, using exactly the same data.  
In each case, they calculate vibrational spectra.  
They find GPR to perform better for a fixed number of data points, with a relatively accurate spectrum from as few as 300 data points.

\subsection{Stability of solids}\label{secStability}

Another important field is the relative stability of different
arrangements of atoms in solids, be they metallic alloys or molecular
crystals.  Searching all possible arrangements is again a Herculean task,
which could be tremendously accelerated if the patterns of the output could
be machine-learned instead of having to be recalculated over and over.

Artrith et~al.~\cite{Artrith} address the problem of creating atomic potentials for alloys.  
There are a few cases where good potentials have been intuited in the past, but the essentially infinite number of possibilities and simulation conditions leads to a strong need for automation.  Essentially, direct simulation with first-principles methods is hopelessly expensive for many problems and properties of interest.  
They use NNs to speed up the sampling for amorphous and disordered materials, and use the subsequent potential to calculate the phase diagram.

On the other hand, Schmidt et~al.~\cite{Schmidt} scan many materials, looking specifically at ternary compounds to find the most stable structures.  
Here they find that ML reduces the calculational cost by about a factor of 4, but the high accuracy needed for such predictions limits the benefits of the ML approach to this problem.

An important problem is that of finding stable polymorphs of molecular crystals.
Li et~al.~\cite{Li} introduce Genarris, a Python package that does inexpensive approximate DFT calculations and analyzes results with a relative coordinate descriptor developed specifically for this task.
It uses ML for \glos{clustering} and can be targeted for various outcomes, ranging from random structure generation to finding a maximally diverse set of structures to seed a genetic algorithm.

A quite different problem is that of grain boundaries in materials, where all sorts of non-stoichiometric defects appear.  
Kiyohara and Mizoguchi~\cite{Kiyohara} use a Monte Carlo tree search to model grain-boundary segregation and test it on silver impurities in copper.  
They find that the search algorithm reduces the number of evaluations by a factor of 100 and yields insight into the nature of the most relevant sites.

Returning to searching chemical compound space, Seko et~al.~\cite{Seko} look at all possible inorganic crystals, which is a much vaster space than those that have been discovered so far.
They propose descriptors to estimate the relevance of chemical composition to stability.
They train and test on experimental databases and also estimate phase stability from first-principles calculations.

Graphene is a promising material for future electronic applications.  
Dieb et~al.~\cite{Dieb} consider doping graphene with boron atoms.  High levels of doping have been recently made and measured.  Their aim is to find the most stable structures, using first principles calculations and ML to perform the search.  
They find useful patterns and predict properties as a function of boron doping.

\subsection{Finding new density functionals}\label{secFunctionals}

Density functional theory (DFT) calculations are currently of limited accuracy and reliability, and
often fail badly for materials that are of key technological interest.
Several of the papers in this Special Topic address the idea of using ML to improve existing functionals or to create entirely new ones.

Mardirossian and Head-Gordon~\cite{Mardirossian} develop ML technology to optimize exchange-correlation functionals at different levels on Jacob's ladder\cite{prtssc2005} of increasing sophistication. 
Their work is at the highest rung, in which a double-hybrid functional is optimized (but not overfitted) to a dataset of nearly 5\,000 molecular energies, screening trillions of possible functionals, but ending up with only 14 parameters. This might prove an invaluable combination of accuracy and computational efficiency.

Another place where ML methods can be fruitfully applied is to find the exact (or at least a much more accurate) exchange-correlation functional, without fitting a given form of approximation.
Nagai et~al.~\cite{Nagai} take small model problems, in which the exact density and energy are known, and use inversion techniques to find the exact Hartree-exchange-correlation energy and potentials.  
In the framework of Levy and Zahariev,~\cite{lz2014} they then train and test a NN for this object.  
This work can be classified as going beyond the existing approximations used currently in DFT.

On the other hand, Ji and Jung\cite{Ji} use a grid-based local representation of various electronic properties to predict DFT energies, densities, and exchange-correlation potentials for 16 small main-group molecules, with errors below 1\,kcal/mol when trained for each molecule separately.  
The errors rise only to 4\,kcal/mol if a small subset of the molecules is used for training, holding out the promise of a transferable method sensitive to chemical environment.

The work of Hollingsworth et~al.~\cite{Hollingsworth} is focused on whether or not simple exact conditions, which have been highly useful in guiding human-based functional design, are useful for improving learning curves of ML functional approximations.
While they examine the question for the Kohn-Sham kinetic energy of simple models, their results should provide a guide for applications to the exchange-correlation energy, such as
in the work of Nagai et~al.~\cite{Nagai}
They find that, while exact conditions do improve learning rates, the improvement is only significant when there is similarity in the densities within the training manifold.

Seino et~al.~\cite{Seino} work with approximate forms for the energy density of the Kohn-Sham kinetic energy to improve over existing approximations to orbital-free DFT.
They expand in higher gradients than are typically included in human approximations, and use ML to find coefficients and density dependencies, and compare their accuracies to many existing orbital-free functionals.

\subsection{Analyzing molecular dynamics simulations}\label{secAnalysis}

Even with classical force fields, there is
tremendous interest in speeding up specific aspects of MD simulations,
such as rare-event sampling or slow, long-term motions of long molecules.
A related interest is the extraction of information from the large amounts of data generated by MD simulations.

The work of Boninsegna et~al.~\cite{Boninsegna} is focused on finding collective variables to determine long-time and coarse-grained motions from MD data.
There is substantial history of ad~hoc intuitive approaches to these
problems, but their Sparse Identification of Nonlinear Dynamics (SINDy) approach does this automatically, and they prove the correctness of their approach in the limit of infinite data.
A similar problem is tackled by Wehmeyer and No{\'e}\cite{Wehmeyer} using a DNN \glos{autoencoder}, which finds low-dimensional features (that is, the slow dynamics of the underlying stochastic processes) embedded in a higher-dimensional feature space.
They test their methodology on simple model systems and a 125\,$\mu$s trajectory of the fast-folding peptide villin.

Finally, Matsunaga and Sugita\cite{Matsunaga} approach this topic from a different viewpoint.  
They construct a Markov state model from MD trajectories and then refine that model using ML methods applied to experimental data.  Thus their methodology attempts to overcome the inherent limitations of the MD force field model by comparison with experiment, whereas the other contributions are focused on speeding up a calculation, but entirely within the MD simulation itself.

\subsection{Everything else}\label{secOther}

Not everything fits into simple categories and that is especially true
in this field, including attempts to improve geometry optimization, to analyze
the statistics behind benchmark datasets, and applications
to larger biopolymers.  In fact, there are many, many more possible
applications of data-enabled chemistry, many of which are not included
in this Special Topic and so are beyond the scope of this editorial.

Pernot and Savin\cite{Pernot} perform an in-depth study of the methods currently being used to benchmark approximations against datasets, an important topic as ever larger datasets are being generated.
They question the summary statistics typically reported, such as RMSE or MAE, showing that because the error distributions are not simple, little can be inferred about error probabilities from these numbers alone.
They advocate more informative measures and show their usefulness.

The position of the LUMO and the width of the optical gap in polymers for solar cells are important for power conversion efficiency. 
J{\o}rgensen et~al.~\cite{Jorgensen} perform first-principles calculations on about 4\,000 monomers and show that a grammar variational \glos{autoencoder} using a simple string representation makes quite accurate predictions, reducing the cost of a search by up to a factor of~5.
Afzal et~al.~\cite{Afzal} model the refraction index of organic polymers by combining first-principles calculations with ML to predict packing fractions of the bulk polymers.

Again, along the lines of solving a material- and property-specific problem, Pilania et~al.~\cite{Pilania} study the effect of lanthanide dopants in inorganic scintillation counter materials.  
They use ML on some key experimentally-measured parameters and combine the results with high-throughput electronic structure calculations to perform screening for materials that exhibit optimized levels of the dopant relative to the gap of the host material.

Another important problem is that of geometry optimization, sometimes at a high level of theory.  Schmitz and Christiansen\cite{Schmitz} use GPR to optimize geometries using numerical gradients.  
They use lower levels of electronic structure calculations, such as Hartree-Fock or MP2, and then calculate differences to higher level theory.  The interpolation introduces errors of no more than microHartrees.

In a similar vein, S{\o}rensen et~al.~\cite{Sorensen} also perform geometry optimization but on materials at an approximate DFT level.  They find that \glos{unsupervised learning} can be used to categorize atoms in many diverse partially ordered surface structures of anatase titanium oxide. They also perform gradient-based minimization of a summed cluster distance resulting from this analysis which allows escape from meta-stable basins and so helps find global minima more quickly.

On the other hand, in a totally different system and regime, Botlani et~al.~\cite{Botlani} use MD to simulate dynamic allostery, in which regulator-induced changes in protein structure are comparable to thermal changes.  
Thus the data must be mined to find patterns in a very high dimensional space to identify mechanisms.  
\glos{Unsupervised} \glos{clustering} shows that regulator binding strongly alters the protein's signalling network, not by changing connections between amino acids as one might naively imagine, but rather by changing the connectivity between clusters.

Antimicrobial peptides interact with simple phospholipid membranes, which is relevant for rational drug design.
Cipcigan et~al.~\cite{Cipcigan} introduce new tools for analyzing the $k$-mer spectrum encoded in antimicrobial databases and ways to visualize membrane binding and permeation of helical peptides.


\section{SUMMARY}

We hope you have found this editorial a useful guide to the important content, the papers in our Special Topic.
We end with some remarks about the nature of the field.  ML has been scoring some impressive successes in various areas of human activity.  There is tremendous hope for similar successes in applications to physical sciences.  
However, progress in this direction requires discovering more subtle rules than in many other arenas.  
So it takes time for researchers to find the best ways to apply ML to their problems.  But practical chemists and materials scientists can now create a dazzling array of different molecular structures and alloys.  Once the progress reported here moves beyond development and proof-of-principle, perhaps we can look forward to new materials and drugs designed with ML methods that build on human intuition but apply it to more possibilities than a human could ever imagine.  We shall see.

\begin{acknowledgments}
\noindent 
K.B.{} acknowledges NSF 1464795.
M.R.{} acknowledges funding from the EU Horizon 2020 program Grant No.~676580, The Novel Materials Discovery (NOMAD) Laboratory, a European Center of Excellence.
O.A.v.L. acknowledges funding from the Swiss National Science Foundation (Nos.~PP00P2\_138932 and 310030\_160067). This work was partly supported by the NCCR MARVEL, funded by the Swiss National Science Foundation.  The guest editors sincerely thank the staff and editors of J. Chem. Phys. for putting this Special Topic together and all the authors for their input into this editorial.
\end{acknowledgments}

\nocite{Hy}

\bibliographystyle{editorial} 
\bibliography{editorial,spec}

\end{document}